\documentclass[traditabstract]{aa}

\usepackage{graphicx}
\usepackage{txfonts}

\def\ms{\,m\,s$^{-1}$}         
\def\kms{\,km\,s$^{-1}$}         
\def\vsini{$v$\,sin\,$i$}      
\def\ms{\hbox{\,m\,s$^{-1}$}}         
\def\m2s2{\hbox{\,m$^{2}$\,s$^{-2}$}} 
\def\kms{\hbox{\,km\,s$^{-1}$}}       
\def\vsini{\hbox{$v$\,sin\,$i_{\star}$}}      
\def\Msun{\hbox{$\mathrm{M}_{\odot}$}}             
\def\Rsun{\hbox{$\mathrm{R}_{\odot}$}}
\def\Mjup{\hbox{$\mathrm{M}_{\rm Jup}$}}
\def\Rjup{\hbox{$\mathrm{R}_{\rm Jup}$}}

\def\teff{T$_{\rm eff}$}
\def\logg{log {\it g}}

\def\mr{$M_\star^{1/3}/R_\star$}
\def\modif{}
\def\modiff{}

\begin{document}
\title{\textit{SOPHIE} velocimetry of \textit{Kepler} transit candidates \thanks{Based on observations made with the 1.93-m telescope at the Observatoire de Haute-Provence (CNRS), France}}
\subtitle{III.  KOI-423b: an 18 {\Mjup} transiting companion around an F7IV star.}

\author{
F. Bouchy \inst{1,2}
\and A.S. Bonomo \inst{3}  
\and A. Santerne \inst{3}
\and C. Moutou \inst{3}
\and M. Deleuil \inst{3}
\and R.~F. D\'iaz \inst{1,2}
\and A. Eggenberger \inst{4}
\and D. Ehrenreich \inst{4}
\and \\
C. Gry \inst{1}
\and T. Guillot \inst{5}
\and M. Havel \inst{5}
\and G. H\'ebrard \inst{1,2}
\and S. Udry \inst{6}
}

\institute{
Institut d'Astrophysique de Paris, UMR7095 CNRS, Universit\'e Pierre \& Marie Curie, 98bis boulevard Arago, 75014 Paris, France
\and Observatoire de Haute-Provence, Universit\'e d'Aix-Marseille \& CNRS, 04670 Saint-Michel l'Observatoire, France
\and Laboratoire d'Astrophysique de Marseille, Universit\'e Aix-Marseille, CNRS, 38 rue Fr\'ed\'eric Joliot-Curie, 13388 Marseille, France
\and UJF-Grenoble 1 / CNRS-INSU, Institut de Plan\'etologie et d'Astrophysique de Grenoble, UMR 5274, Grenoble, 38041, France
\and Universit\'e de Nice-Sophia Antipolis, CNRS UMR 6202, Observatoire de la C\^ote d'Azur, BP 4229, 06304 Nice Cedex 4, France
\and Observatoire de Gen\`eve, Universit\'e de Gen\`eve, 51 Ch. des Maillettes, 1290 Sauverny, Switzerland
}
\date{Received ; accepted }

\offprints{F. Bouchy  \email{bouchy@iap.fr} }

\abstract{ We report the strategy and results of our radial velocity follow-up campaign with the \textit{SOPHIE} spectrograph (1.93-m OHP) of four transiting planetary candidates discovered by the \textit{Kepler} space mission. 
We discuss the selection of the candidates KOI-428, KOI-410, KOI-552, and KOI-423. KOI-428 was established as a hot Jupiter transiting the largest and the most evolved star discovered so far and is described by Santerne et al. (2011a). KOI-410 does not present radial velocity change greater than 120 {\ms}, which allows us to exclude at 3$\sigma$ a transiting companion heavier than 3.4 {\Mjup}. KOI-552b appears to be a transiting low-mass star with a mass ratio of 0.15. KOI-423b is a new transiting companion in the overlapping region between massive planets and brown dwarfs. 
With a radius of 1.22 $\pm$ 0.11 {\Rjup} and a mass of 18.0 $\pm$ 0.92 {\Mjup}, 
KOI-423b is orbiting an F7IV star with a period of 21.0874 $\pm$ 0.0002 days and an eccentricity of 0.12$\pm$0.02. From the four selected \textit{Kepler} candidates, at least three of them have a Jupiter-size transiting companion, but two of them are not in the mass domain of Jupiter-like planets.  {\modif KOI-423b and KOI-522b are members of a growing population of known massive companions orbiting close to an F-type star. This population {\modiff currently appears to be} absent around G-type stars, {\modiff possibly} due to their rapid braking and the engulfment of their companions by tidal decay}.}

\keywords{planetary systems - brown dwarfs - binaries: eclipsing - techniques: photometry - techniques:
  radial velocities - techniques: spectroscopic }

\titlerunning{KOI-423b : an 18 {\Mjup} transiting companion around an F7IV star}
\authorrunning{F. Bouchy et al.}

\maketitle

\section{Introduction}
\label{intro}

Launched in March 2009, \textit{Kepler} is the second space mission designed to find transiting exoplanets by high-accuracy photometry. This mission has already demonstrated its strong capability of detecting transiting exoplanet candidates. Thanks to radial velocity follow-up or/and transit timing variation it has 
established and characterized up to 
eight new transiting planetary systems from the super-Earth regime - \textit{Kepler}-10b (Batalha et al. \cite{batalha2011}) - to the massive and bloated Jupiter regime - \textit{Kepler}-5b (Koch et al. \cite{koch2010}). From the 156,000 stars continuously observed by \textit{Kepler}, 706 exoplanets candidates were first identified and 306 of them were published by Borucki et al. (\cite{borucki2010}) in February 2010.
To establish the planetary nature of these candidates, to assess the fraction of false positives, and 
to characterize the true mass and density of new transiting planets, we started a large program of radial velocity follow-up with the \textit{SOPHIE} high-resolution spectrograph mounted on the 1.93-m telescope at the Observatoire de Haute-Provence Observatory (Bouchy et al. \cite{bouchy2009}, Perruchot et al. \cite{perruchot2008}).   

We present here the strategy and results of our first campaign during third quarter 2010 on 
the four \textit{Kepler} object of interest KOI-428, KOI-423, KOI-552, and KOI-410. Our first target KOI-428 
was revealed as a transiting hot Jupiter with a radius of  1.17$\pm$0.04 {\Rjup} and a mass of 2.2$\pm$0.4 
{\Mjup}  orbiting the largest and the most evolved star (2.13 {\Rsun} and 1.48 {\Msun}) discovered so far with a transiting exoplanet. This planet KOI-428b, which also corresponds to the first \textit{Kepler} planet established from the public data, was described by Santerne et al. (\cite{santerne2011a}).

\section{Selection of \textit{Kepler} candidates}
\label{sect1}

A first set of 306 \textit{Kepler} exoplanet candidates was published by Borucki et al. (\cite{borucki2010}). These candidates are faint stars (14 $\le$ m$_\mathrm{V}$ $\le$ 16) that are not included in the ground-based observation follow-up conducted by the \textit{Kepler} team. 
They were identified from the first 33.5-day segment of the science operations (Q1) from May 13 to June 15, 2009 with a temporal sampling of 29.4 min. 

\begin{table*}
\caption{\textit{Kepler} candidates from Borucki et al. (\cite{borucki2010}) selected for the \textit{SOPHIE} follow-up.}
\begin{tabular}{lllllllll}
\hline
\hline
\textit{Kepler} object   & \textit{Kepler} input catalog & RA & DEC & \textit{Kepler}  & Rp$^a$ & Period & Teff & \textit{SOPHIE} follow-up \\
of interest & & & & mag & [\Rjup] & [days] & [K] & status \\
\hline
KOI-410.01 & KIC-5449777 & 19h28m59.53s   & +40d41'45.8"   & 14.5 & 1.07 & 7.217 & 5968 & planet or {\modif DEB} \\ 
KOI-423.01 & KIC-9478990 & 19h47m50.46s & +46d02'03.5"  & 14.3 & 0.94 & 21.087 & 5992  & brown-dwarf \\ 
KOI-428.01 & KIC-10418224 & 19h47m15.29s & +47d31'35.8" & 14.6 & 1.04 & 6.873 & 6127 & planet \\ 
KOI-552.01 & KIC-5122112 & 19h49m35.58s & +40d13'45.1"  & 14.7 & 1.00 & 3.055 & 6018 & low-mass star \\ 
\hline
\multicolumn{4}{l}{$^a$ Estimated radius of the transiting candidates} 
\end{tabular}
\label{keplerID}      
\end{table*}

From this list of candidates, we defined a selection using the following criteria. 
We first removed all targets fainter than magnitude 14.7, which is close to the magnitude limit of the \textit{SOPHIE} spectrograph. 
One hundred fifteen candidates have a \textit{Kepler} magnitude in the range 13.9 - 14.7. 
We then extracted candidates with an estimated radius greater than  0.8 {\Rjup}. Almost all the known transiting 
planets with $R \ge 0.8$ {\Rjup} have a mass higher than 0.4 {\Mjup}. Lower masses are difficult if not impossible 
to detect with \textit{SOPHIE} in this range of stellar magnitude (Santerne et al. \cite{santerne2011b}).  
We removed the monotransit candidates (with an estimated period longer than 33.5 days).  
This led to a list of 11 candidates. The last step was to try to privilege massive hot-Jupiter and 
brown-dwarf candidates with the aims of better understanding this class of massive objects, of further 
exploring the link between massive planets and brown dwarfs and of deriving the main properties of their parent stars.   
As discussed in Bouchy et al. (\cite{bouchy2011}), almost all transiting massive planets 
and brown dwarfs are found to orbit F-type stars as a consequence 
of rapid rotators, independent of their age (Nordstrom et al. \cite{nordstrom97}).  
We then selected the four hottest stars in our list for our first campaign with an estimated effective temperature greater than 5900 K.  The characteristics 
of these four targets are listed in Table~\ref{keplerID}. 

More recently, Borucki et al. (\cite{borucki2011}) have published an update of the \textit{Kepler} exoplanet candidates from the analysis from the analysis of the second quarter (Q2) data lasting 88.7 days from June 20 to September 16, 2009. We thus downloaded from the MAST 
database\footnote{http://archive.stsci.edu/kepler/data\_search/search.php} these two first quarters
of the publicly available Kepler data for our selected candidates. We then combined our radial 
velocity observations to the analysis of the candidates'  Kepler light curves.

\section{\textit{SOPHIE} observations}

We performed the radial velocity (hereafter RV) follow-up of the four candidates with the \textit{SOPHIE} spectrograph (Bouchy et al. \cite{bouchy2009}, Perruchot et al. \cite{perruchot2008}) installed on the 1.93m telescope in the Observatoire de Haute-Provence, France. \textit{SOPHIE} is a cross-dispersed, high-resolution fiber-fed echelle spectrograph, stabilized in pressure and temperature and calibrated with a thorium-argon lamp. It is mainly dedicated to measure high-accuracy radial velocities on solar-type stars for exoplanets and asteroseismology studies. \textit{SOPHIE} is one of the key facilities in the radial velocity follow-up 
of \textit{CoRoT} and SWASP-North transiting planetary candidates. Data reduction pipeline, observational 
strategy, and analysis tools were developed to optimize the follow-up of transiting candidates.  

Observations were made with the \textit{SOPHIE} high efficiency mode, with a spectral resolution of 39,000 at 550nm, and the slow CCD read-out mode. The obj\_AB observing mode was used, i.e., without acquisition of the simultaneous calibration lamp in order to monitor the background sky on the second fiber. The typical intrinsic stability of \textit{SOPHIE} does not require using the simultaneous calibration, and the sky fiber 
is crucial for removing the scattered moonlight (Santerne et al. \cite{santerne2011b}). 
\textit{SOPHIE} spectra were reduced with the online pipeline. Radial velocities were obtained by computing the weighted cross-correlation function (CCF) of the spectra using a numerical spectral mask corresponding to a G2V star (Baranne et al. \cite{baranne1996}; Pepe et al. \cite{pepe2002}). 

Observations were carried out from 2010 July 15 to 2010 September 13. A total of 28 measurements 
with 1-hour exposure times were done corresponding to an equivalent of  about four nights spread over 24 different nights. 
Such flexibility in the schedule of the observations is crucial to properly cover orbital phases of transiting candidates.

\section{KOI-410}

\begin{figure}[]
\begin{center}
\includegraphics[width=8.5cm]{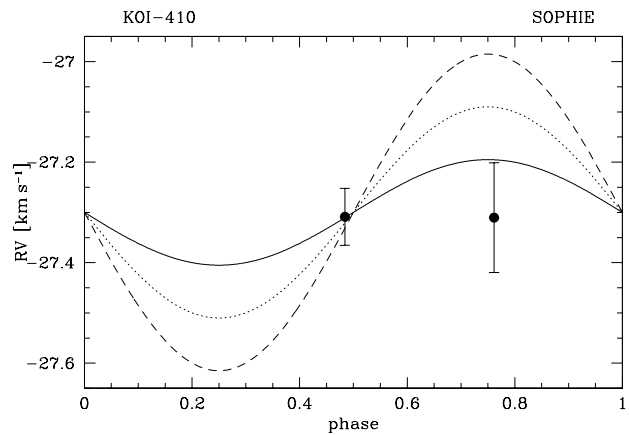}
\caption{Phase-folded \textit{SOPHIE} radial velocity of KOI-410. The three curves correspond to the detection limit at 1, 2, and 3 
$\sigma$.}
\label{fig410}
\end{center}
\end{figure}

\begin{figure}[]
\begin{center}
\includegraphics[width=8.5cm]{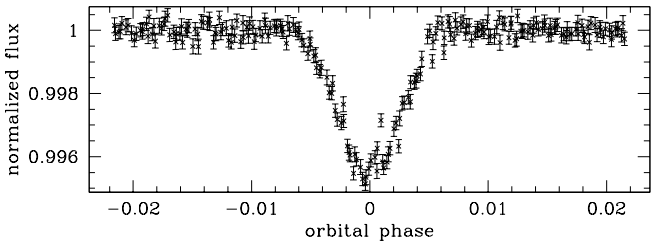}
\caption{Unbinned phase-folded transit light curve of KOI-410.}
\label{transit410}
\end{center}
\end{figure}

We made two \textit{SOPHIE} measurements on this target. Phase-folded radial velocities are shown in Fig.~\ref{fig410} 
and present no significant variation at the level of 120 {\ms}. At 3-$\sigma$ we can exclude a transiting companion with 
mass greater than 3.4 {\Mjup}. The transit light curve, shown in Fig.~\ref{transit410}, is however, significantly V-shaped. 
The most probable contaminating star is KIC-5449780, which 
is an 18.4 \textit{Kepler} magnitude star (36 times fainter) located at 9.6 arcsec. 
{\modif We estimated that, if this star was an eclipsing binary, it would cause a a displacement of the photocenter motion 
in the X and Y directions of more than 5 milli-pixels during the transit, a value excluded by the data (rms of 0.4 milli-pixel 
in both X and Y). This excludes the faint star located at 9.6 arcsec as the origin of the transit.}
{\modif We cannot exclude, however, an unresolved diluted eclipsing binary (DEB).}  
To exclude a transiting giant planet with mass greater than 0.3 {\Mjup} at 3-$\sigma$ requires 
additional RV measurements at the extrema phases with uncertainties of 10 {\ms},  which is beyond the capabilities 
of \textit{SOPHIE} for such a faint star.  

\section{KOI-552}

\begin{figure}[]
\begin{center}
\includegraphics[width=8.5cm]{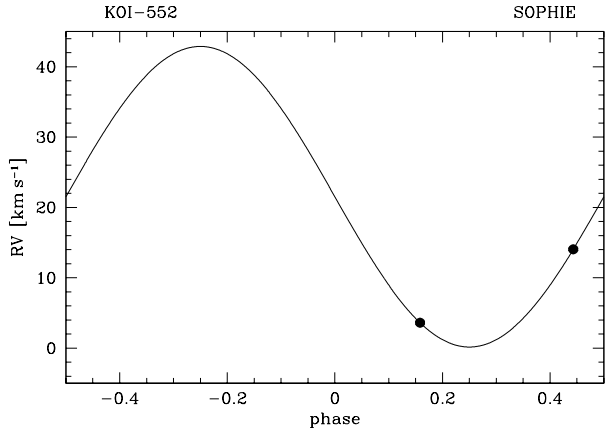}
\caption{Phase-folded \textit{SOPHIE} radial velocity of KOI-552}
\label{rv552}
\end{center}
\end{figure}

We made two \textit{SOPHIE} measurements on this target (see Fig.~\ref{rv552}). They show a large radial-velocity variation in phase with the \textit{Kepler} ephemeris. Assuming 
a circular orbit -- which is reasonable for a three-day period -- the semi-amplitude $K$ is 21.35 $\pm$ 0.14 {\kms} and reveals a low-mass transiting star with a mass ratio of 0.15.

The radius and mass for the central star are estimated to be 1.06 {\Rsun} and 1.10 {\Msun} 
by Borucki et al. (\cite{borucki2011}), respectively. The derived radius of the companion is then 1.0~{\Rjup}.  Such a radius seems small for an M-dwarf companion in this range of mass, where the typical radius is expected 
to be 0.17~{\Rsun}. Borucki et al. (\cite{borucki2011}) recognize that some of the characteristics listed for the stars are uncertain, especially surface gravity. This might introduce errors in stellar diameters up to 25\%. Our two \textit{SOPHIE} spectra
unfortunately have too low signal-to-noise ratio (S/N) to determine the spectral type of KOI-552. 
{\modif From J \& K magnitude and following the same procedure as for KOI-428b ( Santerne et al. \cite{santerne2011a}) and KOI-423b (see section \ref{sectspectra} in this paper), we estimate $T_{\rm eff}$=6560$\pm$150K assuming an [Fe/H]=-0.47 as provided in the KIC catalog, 
a value significantly higher than the estimation given by Borucki et al. (2011), $T_{\rm eff}$=6018K. Assuming a solar metallicity changes the estimated $T_{\rm eff}$ within the uncertainties. If we assume that the effective temperature of the star is close to 6560K, we may expect a larger stellar radius and thus a larger radius for the M-dwarf. We also note that the uncertainties on the impact parameter are large (b=0.79 $\pm$ 0.24), because the transit is V-shape (see Fig.~\ref{transit552} . An underestimation of the impact parameter could also explain the discrepancy in the M-dwarf radius.} 
Our two SOPHIE spectra unfortunately have too low S/N to determine the spectral type of KOI-552.
Additional spectroscopy measurements are required to {\modif better characterize the parameters}  of the detected transiting companion.     

\begin{figure}[]
\begin{center}
\includegraphics[height=8.5cm,angle=90]{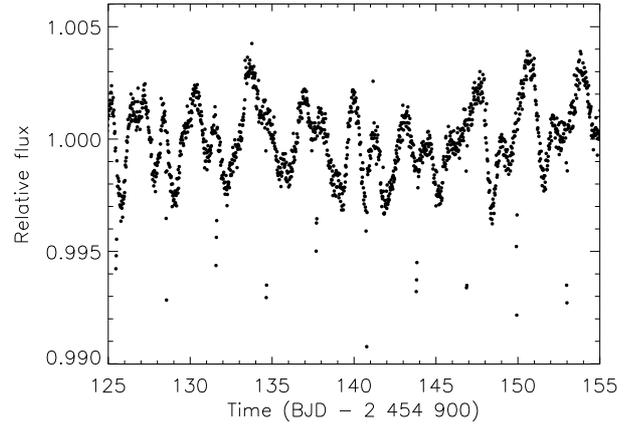}
\caption{{\modif A zoom of the KOI-552 light curve (Q2 data) showing flux variations due to 
starspots and, superposed, transits by a low-mass stellar companion.}}
\label{rot552}
\end{center}
\end{figure}

\begin{figure}[]
\begin{center}
\includegraphics[width=8.5cm]{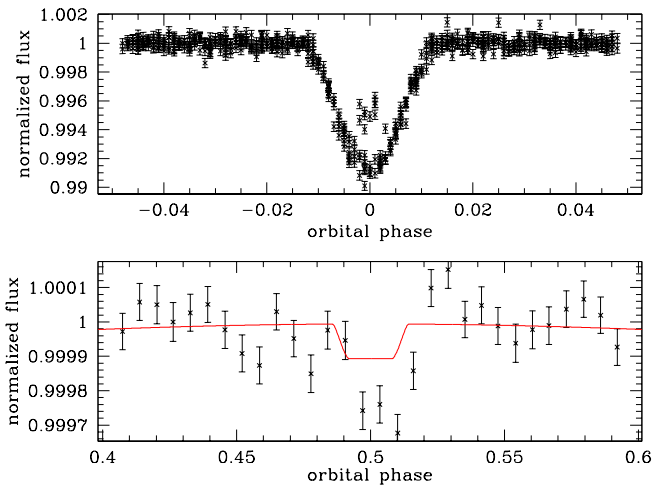}
\caption{Phase-folded light curve showing the unbinned primary transit (upper panel) and the binned secondary transit (lower panel) of KOI-552. The lower pannel also shows the expected model of the secondary transit for a hot Jupiter with a geometric Albedo of 1.}
\label{transit552}
\end{center}
\end{figure}

{\modif From the MAST database, we downloaded the raw Kepler light curve containing photometric data from 
both the first and second quarters. We corrected the Q1 and Q2 light curves separately for long-term trends due 
to systematics and removed a few of the data points clearly affected by instrumental effects, such as those occurring 
just after the two safe mode events in the Q2 light curve (Haas et al. \cite{haas2010}, Jenkins et al. \cite{jenkins2010}). 
We estimated the flux excess due to stellar crowding in the Kepler mask by comparing the values of
the median flux at the beginning of the raw Q1 and Q2 light curves with those of the corresponding error-corrected 
light curves (Jenkins et al. \cite{jenkins2010}). Finally, we subtracted the flux excess from each point
of the Q1 and Q2 raw time series, separately.}
The {\modif treated} light curve presents clear variations due to rotating stellar spots (see Fig.~\ref{rot552}). 
The auto-correlation function of the {\modif light curve was computed after removing the transits, and a clear 
peak was found at $P_{\rm rot}=3.08 \pm 0.15$~days} indicating a possible spin-orbit synchronisation, 
which is expected for this mass ratio. The {\vsini} derived from the CCF is estimated to 17.6 $\pm$ 1.0 {\kms}, 
which agrees with the measured rotational period and a stellar radius 
greater than 1 {\Rsun}. The transit light curve shown in Fig.~\ref{transit552} presents some transits that are shallower  
than others, which may be interpreted by the fact that the transiting companion crosses dark stellar spots. 
The stellar variability with a period equal or close to the orbital period prevents us from properly determining 
the ellipsoidal modulation and the beaming effect (van Kerkwijk et al. \cite{vankerkwijk2010}). 

The \textit{Kepler} Q2 data, released on February 2011 after our spectroscopic observations, permit us to detect the secondary transit (see Fig.~\ref{transit552}). The secondary transit is clearly too deep to be compatible with a perfect reflecting hot Jupiter with the orbital parameters of KOI-552 published in Borucki et al. (\cite{borucki2011}). 
{\modif The occultation depth of $\sim$ 250 ppm would imply a geometric albedo of 2. Assuming a zero albedo, $T_{\rm eff}$=6560 K for the primary star, and radius ratio $R_{c}/R_{\star}$=0.0967, the effective temperature of the companion is estimated to be $T_{\rm eff}$=3300 K, 
which corresponds to an M7 dwarf. Although we cannot exclude the possibility that a fraction of the observed depth might be due to reflected light, 
the secondary transit indicates a low-mass star companion.}  

The V-shape of the transit, the depth of the secondary eclipse, and the possible synchronisation between the companion and the star provide some evidence 
that this candidate was a very unlikely planetary candidate, although it was ranked as priority 2 by the \textit{Kepler} team, . 
Our two RV measurements definitively establish this candidate as a low-mass star.

\section{KOI-423}

\subsection{\textit{Kepler} light curve analysis}

\begin{figure}[h]
\centering
\includegraphics[width=6.5cm, angle=90]{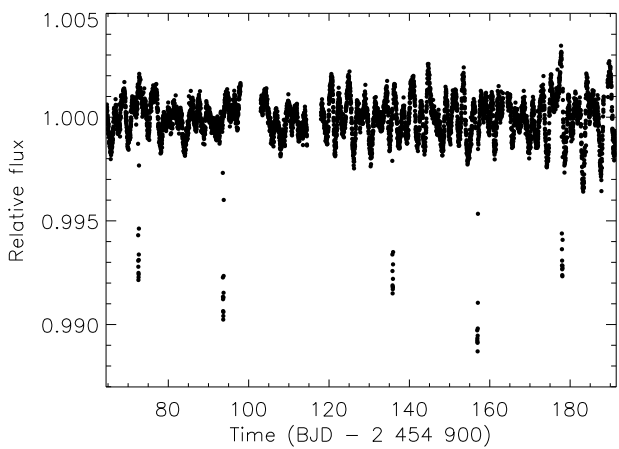}
\caption{The \textit{Kepler} light curve of KOI423 with a temporal sampling of 29.4~min
showing five transits and variations due to stellar activity.}
\label{lcfig}
\end{figure}

The target KOI-423 was observed by \textit{Kepler}  with a temporal sampling of 29.4 min for 122.2 days. 
The publicly-available \textit{Kepler} light curve contains 5708 photometric measurements in total, after 
135 data points were discarded by the \textit{Kepler} pipeline as affected by systematic effects 
(Jenkins et al. \cite{jenkins2010}).
{\modif The raw Q1 and Q2 light curves were downloaded from the MAST and treated in 
the same way as for KOI-552. The obtained Kepler light curve, shown in Fig.~\ref{lcfig}}, exhibits five 
transits with a period of 21.1 days, a depth slightly shallower than 1\% and a duration of about six hours. 
One additional transit should be observable but, unfortunately, it fell entirely inside a gap taking place 
from 2455014.5 to 2455016.7 (BJD). Furthermore, the fourth transit is only partially visible because
of another data gap just before the ingress. 

The KOI-423 light curve shows 
flux variations due to the presence of active regions on the photosphere, 
whose visibility is modulated by the stellar rotation. 
The peak-to-peak amplitude is $\sim 0.6 \%$. 


The standard deviation of the light curve, computed in a 
robust way after removing low frequency variations with a 
sliding median filter, is $2.5\,10^{-4}$, which is compatible
with the median of the errors of the single photometric measurements, 
i.e. $2.1\,10^{-4}$.

To estimate the rotation period of the star, we computed 
the auto-correlation function of the \textit{Kepler} 
light curve, after removing the five transits, {\modif and found a clear 
peak at $P_{\rm rot}=4.35 \pm 0.21$~days.} 
We also computed the Lomb-Scargle periodogram 
of the light curve (Scargle et al. \cite{Scargle82}), and found power at 4.35, 2.17, and 1.45 days 
corresponding to the stellar rotation period and its two first harmonics.

\subsection{\textit{SOPHIE} spectra and radial velocities}
\label{sectspectra}

\begin{figure}[]
\begin{center}
\includegraphics[width=8.5cm]{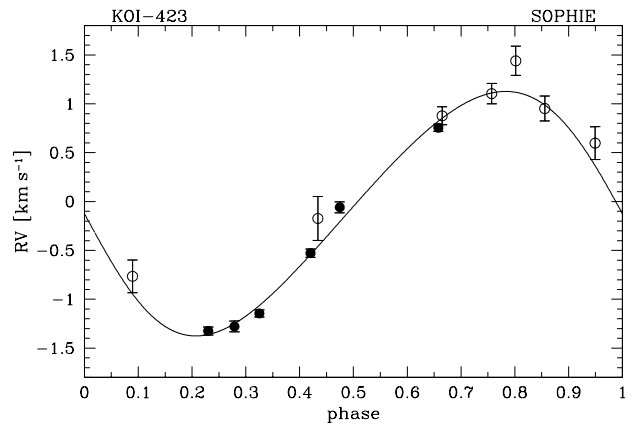}
\caption{Phase-folded radial velocity of KOI-423 with the best Keplerian orbit fit. Open dots correspond to measurements 
affected and corrected by the scattered moonlight.}
\label{rvfig}
\end{center}
\end{figure}

Thirteen measurements were made on this candidate with \textit{SOPHIE} from 2010 July 26 to 2010 
September 10. They are listed in Table~\ref{table423}.  Seven of them were strongly affected by the 
scattered moonlight but corrected for radial velocity analysis using the sky spectra acquired simultaneously. In some exposures, 
the S/N on the sky fiber B, which points on the sky background, was 
even comparable to or greater than on the target fiber A, which is located on the target. We conservatively doubled the uncertainties of these seven measurements and did not use them to compute the bisector span. 

\begin{figure}[]
\begin{center}
\includegraphics[width=8.5cm]{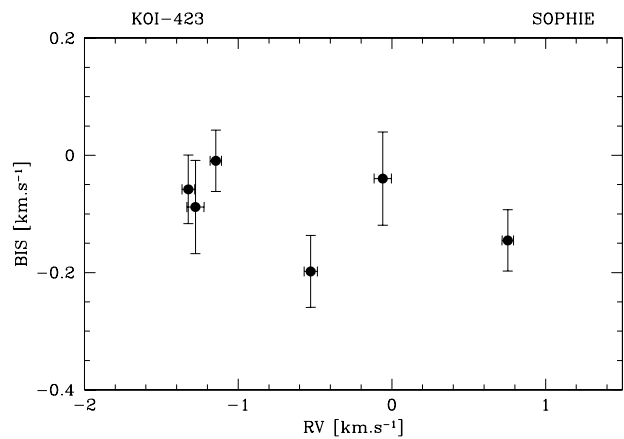}
\caption{Bisector span versus radial velocity of KOI-423 showing no correlation.}
\label{biss423}
\end{center}
\end{figure}

The radial velocities shown in Fig.~\ref{rvfig} present a clear variation in phase with the \textit{Kepler} 
ephemeris and with a slight but significant eccentric orbit.  
The bisector span, listed in Table~\ref{table423}  and shown in Fig.~\ref{biss423}, does not reveal any 
 significant variations at a level more than ten times smaller than the radial velocity variations allowing all 
 scenarios of diluted and blended eclipsing binaries to be excluded. 

We performed the spectroscopic analysis of the parent star using \textit{SOPHIE} spectra. Individual spectra are too poor in quality to allow a proper spectral analysis. 
 The six \textit{SOPHIE} spectra not affected by the scattered moonlight were shifted to the barycentric rest frame and 
 co-added order per order. We finally got a co-added spectrum with an S/N of about 110 
 per pixel on the continuum at 550 nm.  From the analysis of a set of isolated lines, we derived a {\vsini} 
 of 16$\pm$2.5 {\kms}. The spectroscopic analysis was carried out using the same methodology as for the 
 \textit{CoRoT} planets described in detail in Bruntt et al. (\cite{bruntt2010}).

We limited the abundance analysis to the iron ions, because of the high {\vsini} of the star and the quite moderate spectral resolution and S/N of the combined spectrum. In total 53 FeI and 11 FeII spectral lines were used for the temperature and log\,$g$ determination. Following the Bruntt et al. (\cite{bruntt2008}) methodology, abundances were measured differentially with respect to a \textit{SOPHIE} solar spectrum. We found a marked underabundance of iron, with [Fe/H]=-0.29 $\pm$ 0.1. This result was further checked using the \textit{SOPHIE} CCFs. Following the relations established by Boisse et al. (\cite{boisse2010}), we derived a 
{\vsini} of 12.6 $\pm$ 1 {\kms} and an [Fe/H] of -0.18 $\pm$ 0.1 dex. The metallicity is 1-$\sigma$ higher
than the spectroscopic determination. The {\vsini} value estimated from the CCF {\modif is 3-$\sigma$ lower than then spectroscopic determination, which may come from the fact that the relation was calibrated assuming solar metallicity.}
The rotation velocity of KOI-423 derived from the stellar radius and the rotational 
period estimated from the light curve is 16.2 $\pm$ 1.7 {\kms}, in full agreement with the spectroscopic determination, 
indicating that $i_{\star}$ {\modif may} be close to 90 deg. 

The temperature derived from the \ion{Fe}{i} - \ion{Fe}{ii} equilibrium is {\teff} = 6260 $\pm$ 140 K. This is 2 $\sigma$ higher than the value of 5992 K quoted for the star in the Kepler Input Catalog.    
From infrared magnitudes J and K given in the 2MASS catalog, a reddening of E(J-K)=0.091 (from Cardelli et al. \cite{cardelli1989}), and Eq. 3 in Casagrande et al. (\cite{casagrande2010}) assuming [Fe/H] = -0.3 dex, we derived an effective temperature {\teff} = 6380 $\pm$ 150 K. This estimate agrees with our spectroscopic determination.
As already mentioned for KOI-428 (Santerne et al. \cite{santerne2011a}), the infrared determination of the effective temperature from Casagrande et al. (\cite{casagrande2010}), based on stars of luminosity class IV and V, seems 
to be more reliable than the one used for the \textit{Kepler} Input Catalog. 

The surface gravity was estimated using pressure-sensitive lines: the \ion{Mg}{i}1b lines, the \ion{Na}{i} D doublet, and  the \ion{Ca}{i} at 6122 and {\modiff 6162} {\AA}. As already noted by Bruntt et al. (\cite{bruntt2010}), the results obtained on the \ion{Mg}{i} triplet vary from one line to another due to the difficulty in the normalization of these broad lines. Fitting each of the observed lines {\modif of the triplet} with synthetic spectra calculated using MARCS models, we found {\logg} = 3.90 $\pm$  0.25. 
The \ion{Na}{i} D lines gave {\logg} =  4.0 $\pm$ 0.2 and the \ion{Ca}{i} lines {\logg} = 4.1 $\pm$ 0.2. The latter are in good agreement with the value of {\logg} = 4.2 obtained with VWA  {\modif  from the agreement between the \ion{Fe}{i} and \ion{Fe}{ii} abundances. 
Considering the issue on the normalization for broad lines such as \ion{Mg}{i}, we adopted as a final value the gravity derived from the \ion{Ca}{i} pressure sensitive lines, that is, \logg\ = 4.1 $\pm$ 0.20.}

\begin{table}[h]
\centering        
\setlength{\tabcolsep}{1.0mm}
\caption{\textit{SOPHIE} measurements of KOI-423}
\renewcommand{\footnoterule}{}                          
\begin{minipage}[!]{7.5cm}  
\renewcommand{\footnoterule}{}     
\begin{tabular}{lcccc}
\hline
\hline
BJD & RV &   $\sigma_{RV}$ & BIS & S/N/pix \\
-2\,455\,000 & [\kms] & [\kms] & [\kms] & @550nm \\
\hline
403.49308$^a$ & -0.173 &	0.225 &     -  &   8.2 \\
425.43895 & -0.059 &	0.056 &	  -0.040  &  18.2 \\
429.45136$^a$  & 0.877 &	0.092 &   - &  21.1 \\
431.39644$^a$ & 1.104 &	0.104 &   - &  20.3 \\
432.33993$^a$ & 1.440 &	0.150 &   - &  16.4  \\
433.47474$^a$ & 0.952 &	0.127 &   - &  16.5 \\
435.45034$^a$ & 0.598 &	0.168 &   - &  11.9 \\
438.40205$^a$ & -0.765 &	0.167 &   -  &  11.0 \\
441.36981 & -1.324 &	0.041 &	0    -0.058 &  20.7 \\
442.39937 & -1.278 &	0.056 &  0    -0.088 &  18.4  \\
443.37619 & -1.145 &	0.037  &	0    -0.009 &  21.9 \\
445.38001 & -0.528 &	0.043 &	0    -0.198 &  22.7 \\
450.39197 & 0.754 &	0.037 &	0    -0.145 &  23.5  \\
\hline
\end{tabular}
\vspace{-3mm}
\footnotetext[1]{Measurements corrected from the scattered moonlight.}
\label{table423}
\end{minipage}
\end{table}

\subsection{System parameters}

System parameters were derived by performing the transit modeling and
the Keplerian fit of the radial velocity measurements simultaneously. 
Transit fitting was carried out following the formalism 
of Gimenez (\cite{Gimenez06, Gimenez09}) after removing stellar variability 
in the out-of-transit light curve to correctly normalize the transits.
For this purpose, we fitted a third-degree polynomial to the 9h 
intervals of the light curve before the ingress and after the egress of 
each transit. We discarded the fourth transit as only half  
visible, which prevented us from fitting the polynomial before 
the transit ingress and thus normalizing it as for the other ones.

The nine free parameters of our global best fit
are the orbital period $P$, the transit epoch $T_{\rm tr}$,
the transit duration $d_{\rm tr}$,
the ratio of the companion to stellar radii $R_{\rm c}/R_{\star}$,
the inclination $i$ between the orbital plane and the plane of the sky,
the Lagrangian orbital elements $h=e~\sin{\omega}$ and  
$k=e~\cos{\omega}$, where $e$ is the eccentricity and
$\omega$ the argument of the periastron, the radial-velocity 
semi-amplitude $K$, and the systemic velocity $V_{0}$.
The two nonlinear limb-darkening coefficients $u_{+}=u_{a}+u_{b}$
and $u_{-}=u_{a}-u_{b}$~\footnote{$u_{a}$ and $u_{b}$ are the coefficients
of the limb-darkening quadratic law:
$I(\mu)/I(1)=1-u_{a}(1-\mu)-u_{b}(1-\mu)^2$, where $I(1)$ is the 
specific intensity at the center of the disk and $\mu=\cos{\gamma}$, 
with $\gamma$ the angle between the surface normal and the line of sight} 
were fixed in the transit modelling. They cannot be derived as free parameters
because the transit ingress and egress are not sampled well, 
given the coarse temporal sampling of the \textit{Kepler} light curves. 
The adopted limb-darkening coefficients $u_{a}$ and $u_{b}$ for 
the \textit{Kepler} bandpass were taken from Sing's \cite{Sing10} 
tables\footnote{ http://vega.lpl.arizona.edu/singd/David\_Sing/Limb\_Darkening.html}, after 
linearly interpolating at the $T_{\rm eff}$, log\,$g$, and metallicity of the star:
$u_{a}=0.303 \pm 0.014$ and $u_{b}=0.308 \pm 0.005$, which 
give $u_{+}=0.611 \pm 0.015$ and $u_{-}=-0.004 \pm 0.015$. 

The best-fit parameters were found by using the algorithm AMOEBA (Press et al. \cite{Pressetal92}) 
and changing the initial values of the parameters with a Monte-Carlo method 
to find the global minimum of the $\chi^{2}$.
Following Kipping \& Bakos ({\cite{kipping2011}) and Santerne et al. (\cite{santerne2011a}),
the transit modeling was performed with a temporal sampling 
five times denser than that of \textit{Kepler}, i.e. one point every 5.88~min, 
and then binning the model samples to match the \textit{Kepler} sampling rate.
In such a way, the solution of the transit parameters is more accurate 
than a simple $\chi^2$ minimization between the \textit{Kepler} 
measurements and the transit model. Indeed, the \textit{Kepler} sampling of 29.4~min 
inevitably smears all sharp changes occurring, such as  transit
ingress and egress (Kipping \& Bakos \cite{kipping2011}).

\begin{figure}[h!]
\centering
\includegraphics[width=5.5cm, angle=90]{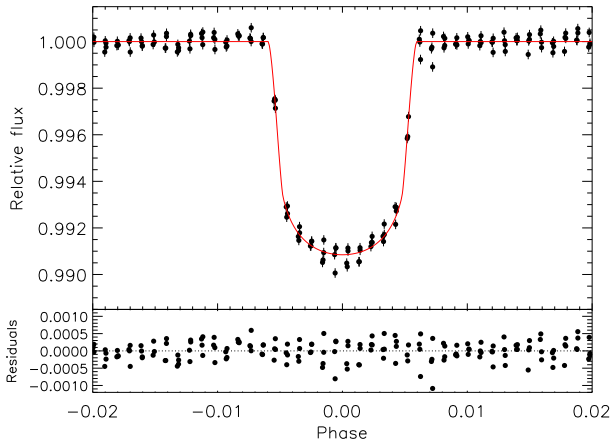}
\vspace{+0.5cm}
\caption{
\emph{Top panels}: Unbinned phase-folded \textit{Kepler} transit light curve of KOI-423.
The red solid line shows our oversampled 
transit model. 
\emph{Bottom panels}: the residuals from the best-fit model.}
\label{tr_bestfit_fig}
\end{figure}

Fitted and derived system parameters are listed in
Table~\ref{systemparam}, together with their 1-$\sigma$ errors
estimated using a bootstrap procedure. The last works as follows:
1) subtracts the best solution to the data;
2) shifts the photometric residuals, thus taking also possible 
correlated noise into account; 3) shuffles a fraction of the radial velocity residuals 
(typically $ \rm 1/e\simeq37 \%$); 4) adds the subtracted 
solution; and 5) again carries out the global best fit. 
The limb-darkening parameters $u_{+}$ and $u_{-}$ in the transit modeling 
were allowed to vary within their error bars related to the 
atmospheric parameter uncertainties. 
More than one thousand iterations were used for this procedure.
Figure~\ref{rvfig} shows the phase-folded radial velocity of KOI-423 with the best Keplerian fit.
Figure~\ref{tr_bestfit_fig} shows the phase-folded transit light curve of KOI-423 and, superposed, 
the transit model with a sampling rate of $5.88$ min.

\begin{table}
\centering
\caption{KOI-423 system parameters.}            
\begin{minipage}[!]{7.0cm} 
\renewcommand{\footnoterule}{}                          
\begin{tabular}{l l}        
\hline\hline                 
Parameters   & Value \\
\hline
Transit epoch $T_{tr}$ [BJD] & $54972.5959_{-0.0005}^{+0.0006}$  \\
Orbital period $P$ [days] & 21.0874 $\pm$ 0.0002 \\
Transit duration $T_{\rm 14}$ [h] & 6.02 $\pm$ 0.09 \\
Radius ratio $R_{c}/R_{\star}$ &  $0.0896_{-0.0012}^{+0.0011}$ \\
Inclination $i$ [deg] & $88.83_{-0.40}^{+0.59}$ \\
$h=e~\sin{\omega}$ &  $0.120_{-0.024}^{+0.022}$ \\ 
$k=e~\cos{\omega}$ &  $-0.019_{-0.012}^{+0.015}$ \\ 
Semi-amplitude $K$ [\kms] & $1.251_{-0.027}^{+0.030}$ \\
Systemic velocity  $V_{0}$ [\kms] &  $-0.101_{-0.015}^{+0.017}$ \\
\\
Orbital eccentricity $e$  &  $0.121_{-0.023}^{+0.022}$ \\
Periastron argument $\omega$ [deg] & $98.9_{-6.8}^{+5.9}$ \\ 
Scaled semi-major axis $a/R_{\star}$ & $23.8_{-1.7}^{+1.8}$ \\
{\mr} [solar units] & $0.740_{-0.051}^{+0.057}$ \\
Stellar density $\rho_{\star}$ [$g\;cm^{-3}$] & $0.57_{-0.11}^{+0.14}$ \\
Impact parameter $b$ & $0.43_{-0.18}^{+0.11}$ \\
\\
Effective temperature $T_{eff}$[K] & 6260$\pm$140 \\
Surface gravity log\,$g$ [dex]&  4.1$\pm$0.2  \\
Metallicity $[\rm{Fe/H}]$ [dex]&  -0.29$\pm$0.10 \\
Rotational velocity {\vsini} [\kms] & 16$\pm$2.5 \\
Spectral type & F7IV \\
\\
Star mass [\Msun] &  $1.10_{-0.06}^{+0.07}$ \\   
Star radius [\Rsun] &  1.39$^{+0.11}_{-0.10}$  \\
Distance of the system [pc] & 1200$\pm$250  \\
\\
Orbital semi-major axis $a$ [AU] & 0.155 $\pm$ 0.003\\
Companion mass $M_{c}$ [\Mjup] &   $18.00_{-0.91}^{+0.93}$ \\
Companion radius $R_{c}$[\Rjup]  &  $1.22_{-0.10}^{+0.12}$   \\
Companion density $\rho_{c}$ [$g\;cm^{-3}$] &  $12.40_{-2.6}^{+3.4}$ \\
Equilibrium temperature  $T_{\rm eq}$ [K]~$^a$ & $905_{-37}^{+39}$  \\
\hline       
\multicolumn{2}{l}{$^a$  black body equilibrium temperature for an isotropic}\\
\multicolumn{2}{l}{ planetary emission.}
\end{tabular}
\end{minipage}
\label{systemparam}     
\end{table}

The star's mass and radius were estimated using a grid of STAREVOL evolutionary stellar tracks calculated for the \textit{CoRoT} exoplanet program (Palacios, 2007, {\sl private com}). We used the regular approach for exoplanet host stars, which is the comparison of the star location in a ({\teff}, {\mr}) HR-plane to stellar evolutionary tracks with the appropriate range of metallicity. Using the \mr\ from the transit modeling (see Table~\ref{systemparam}) and the effective temperature, we estimated the mass M$_\star$ = 1.10  $^{+0.07}_{-0.06}$ from the tracks and deduced the radius from this mass value and the \mr values. Combining these mass and radius values we get {\logg} = 4.20 $\pm$ 0.2, which is in good agreement with the spectroscopic value. With the adopted stellar parameters, we derived 
for the transiting companion M$_{c}$ = 18.0 $\pm$ 0.92 {\Mjup} and R$_{c}$ = 1.22 $\pm$ 0.11 {\Rjup}.  

The distance of the star can be estimated from the V magnitude (mv = 14.46), a bolometric correction BC= -0.03 
for F7IV stars (Straizys \& Kuriliene \cite{straizys1981}),  the adapted solar bolometric magnitude ($M_{\rm bol,\odot}$=4.74) 
recommended by Torres (\cite{torres2010}), the reddening coefficient A$_{\rm V}$=0.423 provided by MAST database, and the stellar luminosity derived from the stellar radius and {\teff}. We find $d$= 1200$\pm$250 pc. 
{\modif The age was estimated to be 5.1$\pm$1.5 Gyr and 4.4$\pm$1.0 Gyr from the STAREVOL and the CESAM stellar evolution models, respectively.}  All the derived stellar parameters are reported in Table~\ref{systemparam}.

\subsection{KOI-423b and the evolution of brown dwarfs and massive exoplanets}
\label{sec64}

\begin{figure}[]
\begin{center}
\includegraphics[width=8.5cm]{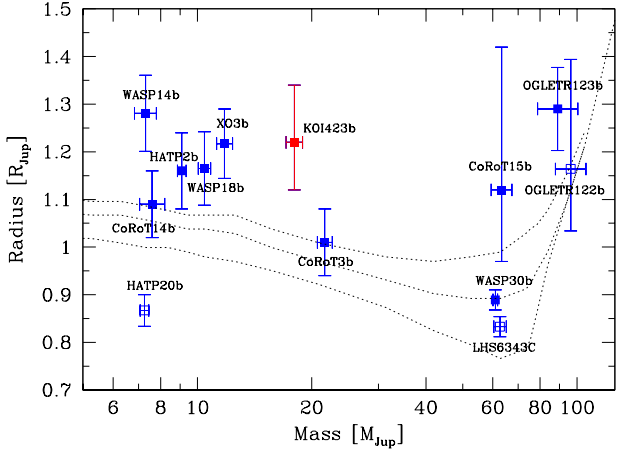}
\caption{Mass-radius diagram of massive transiting planets, eclipsing brown-dwarfs, and low-mass stars. 
Open symbols correspond to host stars with {\teff} $\le$ 6000 K. Dotted lines correspond to theoretical isochrones 
with ages 0.5, 1, and 5 Gyr (from top to bottom) from Baraffe et al. (\cite{baraffe2003}).}
\label{plotmr}
\end{center}
\end{figure}

\begin{figure}[]
\begin{center}
\includegraphics[width=8.5cm]{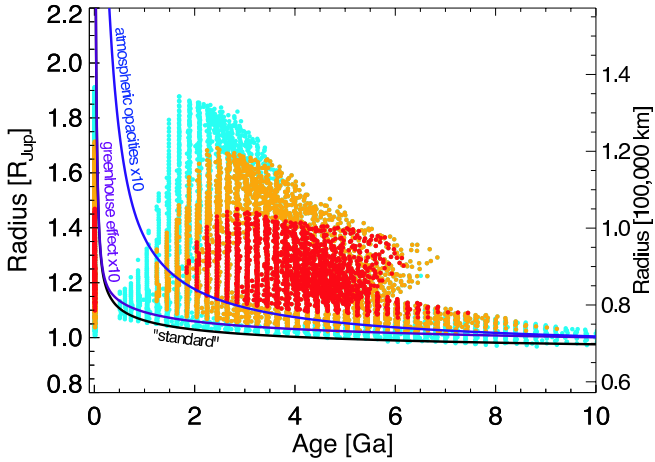}
\caption{Derived constraints on the radius of KOI-423b as a function of age compared to evolution models. The dots represent solutions within $1\sigma$ (red), $2\sigma$ (orange), and $3\sigma$ of the inferred $(T_{\rm eff},\rho_\star)$ obtained from stellar evolution models. From bottom to top, the lines correspond to evolution calculations for a hydrogen-helium brown dwarf with $Y=0.25$ using a ``standard" evolution model (black line), a model in which the greenhouse factor has been arbitrarily multiplied by 10 (purple line), and a model in which the atmospheric opacities have been arbitrarily multiplied by 10 (blue line). }
\label{plotrage}
\end{center}
\end{figure}

Figure~\ref{plotmr} compares the size of KOI-423b to other known transiting objects from 5 {\Mjup} to the M-dwarf mass range. Its size is large, and it appears to be inflated, although comparable to other massive planets such as XO-3b (Johns-Krull et al. \cite{johns2008}). However, given its high mass, it lies farther from ``classical'' evolution tracks, such as those for isolated objects from Baraffe et al. (\cite{baraffe2003}), than other known objects of this class. The problem is made even more acute when one considers that KOI-423b is relatively cold ($T_{\rm eq}=905$\,K), in contrast to most other objects in this plot, implying that irradiation effects cannot be held responsible for its anomalously large size. 

In Fig.~\ref{plotrage} we compare the constraints in age and radius obtained from photometric, spectroscopic, and CESAM stellar evolution models to dedicated evolution models including irradiation (see Guillot \& Havel \cite{guillot2011} and Havel et al. \cite{havel2011} for a description of the method). We first calculate a ``standard'' evolution track with a simplified semi-gray atmospheric boundary condition that has been parametrized to agree with radiative transfer calculations and evolution calculations (see Guillot \cite{guillot2010}). Specifically, we adopt a thermal (infrared) opacity coefficient $\kappa_{\rm th}=0.04\rm\,cm^2\,g^{-1}$ and a visible opacity $\kappa_{\rm v}=0.024\rm\,cm^2\,g^{-1}$. The corresponding evolution track is found to be at the $3\sigma$ lower limit of the main-sequence solutions. When we increase the greenhouse effect by a factor 10 by decreasing the visible opacity by the same factor, the consequence on the evolution is small. This is because the deep atmosphere is convective and weakly affected by a deeper penetration of the stellar light. Similarly, recipes commonly adopted to explain the large size of exoplanets of smaller sizes, including tides, dissipation of kinetic energy, or increased interior opacities (see Guillot 2008) have little effect on this very large planet. Even the possibility that only 50\% of the atmosphere would be allowed to transfer heat efficiently (proposed by Bouchy et al. \cite{bouchy2011} as one possibility to explain the large size of CoRoT-15b) does not work in this case. 
{\modiff We also checked that the energy dissipation from the circularization of the orbit by tides is quite inefficient at the distance 
of KOI-423b.To obtain the required dissipation rates, one would need a tidal quality factor Q$_p$ of about 10, i.e., at least 3 orders of magnitude over Jupiter's.} In fact, the only way we found to reconcile models and observations within $\sim 1\sigma$ is to invoke an ad-hoc increase in the atmospheric opacities, by an order of magnitude or more (we used $\kappa_{\rm th}=0.4 \rm\,cm^2\,g^{-1}$ and $\kappa_{\rm v}=0.24\rm\,cm^2\,g^{-1}$). This is in essence similar to what is proposed by Burrows et al. (\cite{burrows2007}) for planets with lower masses. This yields a slower cooling of the object that remains large for a longer time. 

However, the comparison between KOI-423b and CoRoT-3b (Deleuil et al. \cite{deleuil2008}) adds to the puzzle: the two companions have different sizes (CoRoT-3b lies where theory predicts it should be, KOI-423b is significantly larger), but they otherwise have very similar characteristics (mass,  characteristics of the parent star). KOI-423 has $v\sin i=16\pm2.5\rm\,km\,s^{-1}$, whereas CoRoT-3 has $v\sin i=17\pm1\rm\,km\,s^{-1}$, showing that both are relatively fast rotators. CoRoT-3 has a metallicity close to the Sun ($\rm [M/H]=-0.02\pm 0.06$) when KOI-423b is metal-poor. The smallest massive planet
HAT-P-20b (Bakos et al. \cite{bakos2010}) orbits a host star with the strongest metallicity index [Fe/H] = 0.35.  
On one hand, this follows the trend observed at lower masses that planets around metal-rich stars contain more heavy elements and are therefore on average smaller (see Guillot \cite{guillot2008}, Bouchy et al. \cite{bouchy2010}). On the other hand, one would expect that the atmosphere of KOI-423b is poorer in heavy elements, hence yielding smaller opacities, and potentially, a smaller size. 

One natural possibility would be to invoke age as a way to distinguish between the two objects. However, given the estimated relatively high age, it is difficult to imagine that KOI-423 would be on the pre-main sequence (see Fig.~\ref{plotrage}), whereas CoRoT-3b would be much older. We cannot pretend that we understand the large radius of KOI-423.

\subsection{Comparison to the population of massive exoplanets}

\begin{figure}[]
\begin{center}
\includegraphics[width=9cm]{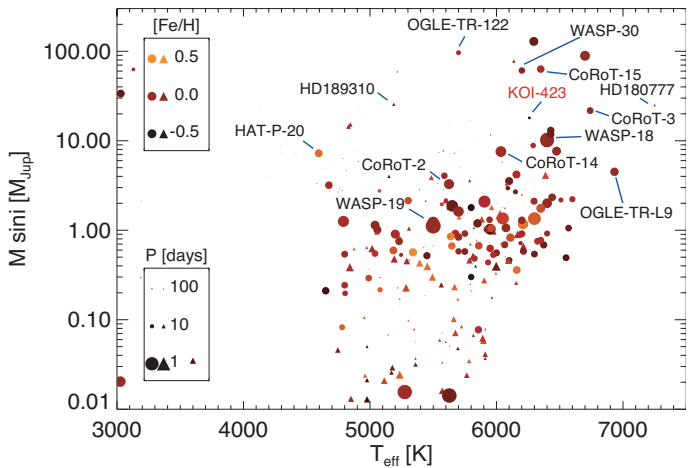}
\caption{Values of the stellar effective temperature and mass of the companion for systems containing a companion in the planet, brown dwarf, or M-dwarf mass range. Transiting/eclipsing systems are shown by circles, and systems only detected in radial velocimetry are shown by trianges (in which case the mass value corresponds to the mass times the sine of the inclination of the orbit with respect to the observer). The size (area) of the symbols is inversely proportional to the orbital period (see labels). The metallicity of the parent star is shown by the symbol color (see labels). Some objects deemed important for understanding the link between these parameters are labeled. The error bars on the masses are small ($\sim 10\%$) and would be barely visible in this diagram. The error bars on the effective temperatures are believed to range from $\sim 100$\,K to up to $\sim 300$\,K (for example for the faint OGLE-TR-122).  
}
\label{plottemp}
\end{center}
\end{figure}

For these systems with a known substellar companion, we plot  in Fig.~\ref{plottemp} the stellar effective temperature as a function of the companion mass for transiting objects or $M\sin i$ value for nontransiting objects. The data for low-mass companions ($<~13\rm M_{Jup}$) was obtained from exoplanets.org (see Wright et al. \cite{wright2011}) and exoplanet.eu. For objects with higher masses, no publicly available database exists, and we manually extracted parameters for the following transiting substellar systems:  OGLE-TR-106 (Pont et al. \cite{pont2005a}), OGLE-TR-122 (Pont et al. \cite{pont2005b}), OGLE-TR-123 (Pont et al. \cite{pont2006}), HAT-TR-205-013 (Beatty et al. \cite{beatty2007}), CoRoT-15 (Bouchy et al. \cite{bouchy2011}), WASP-30  (Anderson et al. \cite{anderson2011}), and LHS6343C (Johnson et al. \cite{johnson2011}). 
And we did it for non-transiting short-period ($P<30$ days) substellar systems: NLTT41135C (Irwin et al. \cite{irwin2010}), HD189310 (Sahlmann et al. \cite{sahlmann2011}), and HD180777 (Galland et al. \cite{galland2006}).

{\modif Although our ensemble of objects is clearly incomplete and derived from inhomogeneous databases, the lack of short period ($P~<10$ days), massive ($M_c >5\rm\,M_{Jup}$) companions around G-type stars ($T_{\rm eff}$ between 5000 and 6000K) appears significant and puzzling. In comparison, these massive companions are routinely found next to F-type stars (including KOI-423). Given that selection biases would tend to favor their detection around G dwarfs, we believe that their absence is real. 

The empty region is bounded very approximatively by the following objects: HAT-P-20b, a massive $7.5\,\Mjup$, anomalously small (see Fig.~\ref{plotmr}) planet orbiting a K7 dwarf in 2.88 days; CoRoT-2b, an anomalously large $3.3\,\Mjup$ planet orbiting a G7 dwarf in 1.74 days; CoRoT-14b, a $7.6\,\Mjup$ planet orbiting an F9V star in 1.5 days; KOI-423b; WASP-30b, a $60\,\Mjup$ brown-dwarf orbiting a relatively young (1-2 Ga) F8V star. At the top of the diagram, OGLE-TR-122b is a $96\,\Mjup$ M dwarf on a 7.27-day orbit around an apparent G dwarf. However, in this case, given the faintness of the star and its poorly-determined $T_{\rm eff}$ ($5700\pm 300\,$K), it may in fact also belong to the F-dwarf category. 

Although this is not shown in Fig.~\ref{plottemp}, we believe that this trend continues well into the M-dwarf regime. For example, a study of 18 OGLE transit candidates with periods between 0.8 and 13.9 days (Bouchy et al. \cite{bouchy2005}) identified two planets (OGLE-TR-10b and OGLE-TR-56b; present in Fig.~\ref{plottemp}), 7 binaries consisting of an M-dwarf orbiting an F-dwarf and 1 M-dwarf around G-dwarf binary. 

It is unlikely that massive companions do not form close to G-dwarfs or do not migrate there: the difference in mass between G-dwarfs and F-dwarfs (~20\%) is small compared to the extension in mass of the ``G-dwarf companions desert'' (from about 10$\Mjup$ to well within the stellar-mass regime). Following Bouchy et al (\cite{bouchy2011}), we propose that close-in massive planets, brown dwarfs, or M-dwarfs may survive when orbiting orbiting close to F-dwarfs, but not when close to G-dwarfs. This could be because around G-dwarfs, the star would rapidly spin down and tidal interactions would lead to a migration and eventual engulfment of its companion. In contrast, around F-dwarfs, because of their small or absent convective zone, the much weaker braking (see e.g. Barker \& Ogilvie \cite{barker09}) will avoid the loss of angular momentum and the rapid decay of the companion's orbit. 


Further work is required including both a proper statistical analysis of the observed correlations and models that combine the formation context, tidal, and magnetic interactions between the companions. {\modiff Kepler will certainly help in providing a 
much more well-sampled comparison between F and G-dwarfs giant planetary companions.}  
}

\section{Conclusions}

The \textit{SOPHIE} follow-up of four \textit{Kepler} transiting candidates, selected with the aim of privileging massive planets and 
brown dwarfs, revealed at least three real massive transiting companions with Jupiter sizes: KOI-428b, a 2.2 {\Mjup} transiting 
planet (Santerne et al. \cite{santerne2011a}); KOI-552b, a $\sim$ 0.17 {\Msun} eclipsing low-mass star; and KOI-423b, a 18.0 {\Mjup} 
transiting companion. The status of the fourth candidate KOI-410 has not yet been solved, but we can exclude a companion 
heavier than 3.4 {\Mjup}, and we still may suspect a possible background eclipsing binary. Our result shows that at least three \textit{Kepler} candidates over the four selected are real Jupiter-size transiting objects, but only one of them, KOI-428b is in the mass domain of Jupiter-like planet.    

KOI-423b is one of the rare types of transiting companions, like \textit{CoRoT}-3b (Deleuil et al., \cite{deleuil2008}) that is located in the overlapping region between massive planet and brown dwarf and which may be considered 
either as a ``superplanet'' or a low-mass brown dwarf. As discussed in section \ref{sec64},  its anomalous bloated radius is clearly not understood. Additional high S/N and high-resolution spectroscopic observations, 
as well as additional \textit{Kepler} data, will permit us to better constrain the size of KOI-423b. 

From RV surveys, only five massive planets or brown dwarfs (M$_c$.sin\,$i >$ 5 {\Mjup}) with short periods 
(P $<$ 30 days) were found including HD162020b (14.4 {\Mjup}), HD41004Bb (18.4 {\Mjup}), HD180777b 
(25 {\Mjup}), HD189310b (25.6 {\Mjup}), and NLTT41135C (33.7 {\Mjup}) respectively orbiting K2V, M2V, A9V, K2V, and M5V stars. Except for HD180777b found in a dedicated RV surveys around A-F type star (Galland et al. \cite{galland2006}), this lack of short-period massive planets and brown dwarfs probably comes from the bias in the selection of RV surveys focusing on spectral types later than F5 and on slow rotators. 

After our first \textit{SOPHIE} campaign, Borucki et al. (\cite{borucki2011}) published an update of the \textit{Kepler} exoplanet 
candidates totalling 1235 candidates. Using the same criteria of selection as presented in section~\ref{sect1} 
(R $\ge$ 0.8 {\Rjup}, mv $\le$ 14.7 and {\teff} $\ge$ 5900 K), we found 17 additional candidates\footnote{{\modiff This list includes the recently announced Kepler-14b (Bucchave et al. \cite{bucchave2011}), a close-in massive planets of 8.4 {\Mjup} orbiting close 
to an F-type star.}}. Radial velocity follow-up of these candidates may help to increase the statistic, and to further reveal the properties 
of massive planets and brown dwarfs, as well as to assess the true fraction of false positives in order to correctly interpret the \textit{Kepler} candidate population.

\begin{acknowledgements}
We thank the technical team at the Observatoire de Haute-Provence for their support with the \textit{SOPHIE} instrument and the 1.93-m telescope and, in particular, for the essential work of the night assistants. Financial support from the ÒProgramme national de plan\'etologieÓ (PNP) of the CNRS/INSU, France, and from the Swiss National Science Foundation (FNSRS) are gratefully acknowledged. We also acknowledge support from the French National Research Agency (ANR-08-JCJC-0102-01). ASB is supported by a CNES grant. AE is supported by a fellowship for advanced researchers from the Swiss National Science Foundation.
\end{acknowledgements}

\end{document}